# Self-aligned patterning technique for fabricating high-performance diamond sensor arrays with nanoscale precision

Mengqi Wang[1,2,3], Haoyu Sun[1,2,3], Xiangyu Ye[1,2,3], Pei Yu[1,2,3], Hangyu Liu[1,2,3], Jingwei Zhou[1,2,3], Pengfei Wang[1,2,3], Fazhan Shi[1,2,3], Ya Wang[1,2,3*] and Jiangfeng Du[1,2,3*]

[1] *Hefei National Laboratory for Physical Sciences at the Microscale and School of Physical Sciences, University of Science and Technology of China, Hefei 230026, China*

[2] *CAS Key Laboratory of Microscale Magnetic Resonance, University of Science and Technology of China, Hefei 230026, China*

[3] *CAS Center for Excellence in Quantum Information and Quantum Physics, University of Science and Technology of China, Hefei 230026, China*

*Authors to whom correspondence should be addressed: ywustc@ustc.edu.cn and djf@ustc.edu.cn*

## abstract：

To efficiently align the creation of defect center with photonics structure in nanoscale precision is one of the outstanding challenges for realizing high-performance photonic devices and the application in quantum technology such as quantum sensing, scalable quantum systems, and quantum computing network. Here, we propose a facile self-aligned patterning technique wholly based on conventional engineering technology, with the doping precision can reach ~15nm. Specifically, we demonstrate this technique by fabricating diamond nanopillar sensor arrays, which show high consistency and near-optimal photon counts, high yield approaching the theoretical limit, and high filtering efficiency for different NV centers. Combined with appropriate crystal orientation, a saturated fluorescence rate of 4.65Mcps and the best reported fluorescence-dependent detection sensitivity of $1900 \text{ cps}^{-1/2}$ are achieved. This technique applicable to all similar solid-state systems should facilitate the development of parallel quantum sensing and scalable information processing.

## Introduction

Arranging artificial atom qubits in diamond to formulate spin arrays with long coherence times[1] and efficient spin-photon interface[2], is attractive for parallel quantum sensing and scalable information processing. Although individual spin qubits integrated with photonic structures[3-16] can now achieve excellent performance including functional quantum repeater nodes[17-19] and nanoscale magnetic field sensors[20-22], placing emitters into the optimal position of the individual photonic nanostructures remains an outstanding challenge[23]. In addition to achieving nanoscale doping [24-26], the alignment of doping to the photonic structure also requires nanoscale precision, which is challenging and a critical barrier to large-scale quantum information processing with diamond qubits.

Enthusiastic efforts are thus devoted in the past few years to achieving above goal by combining and aligning two independent fabrication systems of the diamond doping and photonic structure etching, respectively. One conventional strategy is to fabricate the photonic structure at the pre-located position of color centers determined by optical fluorescence imaging[9]. Due to the randomness of color centers, the alignment process of this method is complex and inefficient. Significantly, the engineering of large-scale photonic device arrays is challenging to achieve using this strategy. Another one is to target-implanted ions into photonic structures prepared in advance. It thus requires a specially designed implantation system, like Atomic force microscope (AFM) assisted ion implantation[27-29] and focused ion beam (FIB) implantation[30-35]. The implantation precision is limited by the size of the mask hole or the ion beam spot, and the alignment accuracy is determined by how precisely the structure is imaged. The reported pre-located or implantation precision and alignment accuracy are 20-30 nm[9,34,36]. A promising alternative strategy to alleviate these constraints is self-aligned patterning, which combines the independent processes through a single pattern, thereby eliminating the aligned inaccuracy and greatly simplifying the design and fabrication process, and has been widely used in semiconductor chip manufacturing. Recently preliminary demonstrations of this strategy in diamond nanofabrication have been implemented [37,38], but the manual high-precision hard mask transfer process [39] is required, limiting its scalability. More importantly, the expected advantage of self-alignment in making high-performance diamond devices is not yet demonstrated.

Here, we introduce a facile self-aligned patterning technique wholly based on conventional engineering technology, and the doping precision of ~15nm is achieved. It applies to various photonic structure fabrication based on circular mask etching, like nanopillar, parabolic reflector, planar waveguides, and ring resonator, etc. Specifically, we demonstrate this technique by fabricating diamond nanopillar sensors[6,7,40], which are widely used in nanoscale magnetic resonance spectroscopy[41] and scanning imaging[42-47]. The self-aligned sensor arrays show high consistency and near-optimal photon counts, high yield approaching the theoretical limit, and high efficiency for distinguishing the number of NV centers. We further controlled the emission dipole orientation through the diamond crystal orientation, achieving a saturated fluorescence rate of 4.65Mcps and the best reported fluorescence-dependent detection sensitivity.

## Results

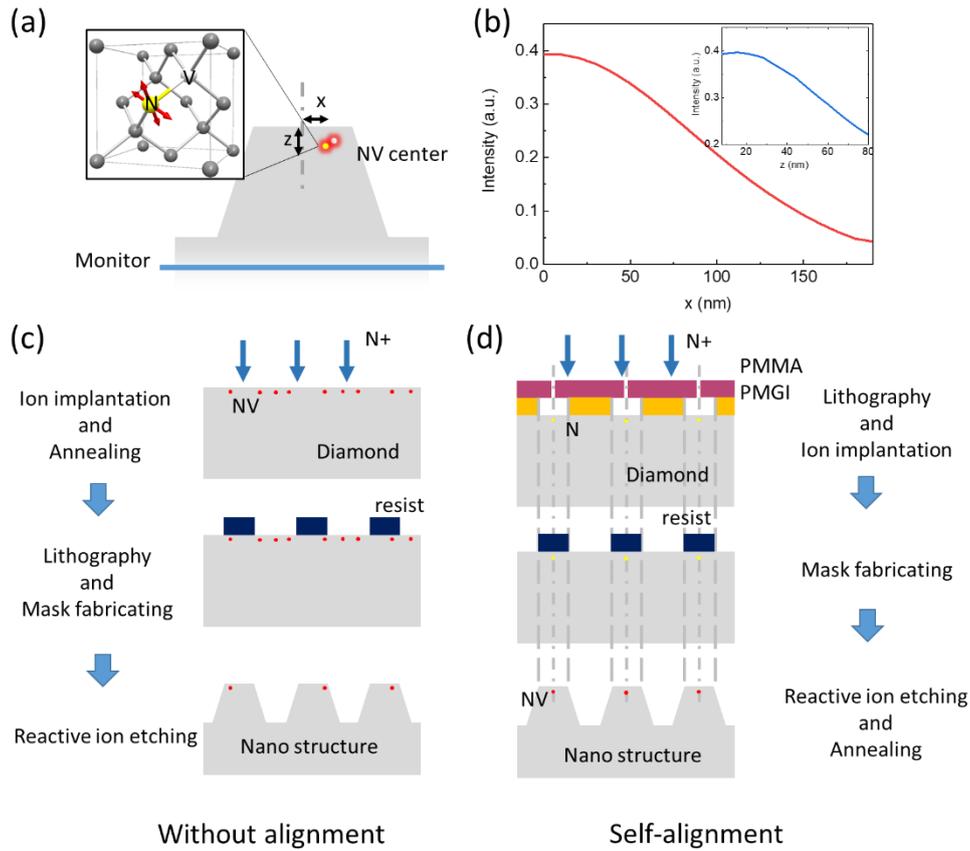

Figure 1 The conceptual demonstration the difference of fabricating diamond nanopillar sensors with or without self-aligned technique. (a, b) The FDTD simulation model (see methods) and results illustrating the importance of precise localization of NV centers into nano-pillar sensors. The fluorescence intensity collected by objective(NA=0.7) as a function of x(z=8nm). The inset shows the fluorescence intensity as a function of z(x=0) (c) Illustration of the fabrication process of photonic structures based on top-down method without alignment. The NV center is generated by maskless ion implantation and annealing. The resist is prepared by photolithography or electron beam lithography, and the photonic structure is formed by reactive plasma etching. (d) Illustration of the fabrication process of photonic structures based on self-alignment. PMGI+PMMA double-layer on diamond is used to constrain the position of the etching mask and ion implantation region.

**The fabrication process based on self-aligned patterning.** Fig. 1 (a) shows the diamond nanopillar sensor with a single NV center embedded in it. Fig. 1 (b) shows the simulated dependence of fluorescence intensity collected from the monitor side in both longitudinal and lateral directions. For a typical top diameter of 380 nm, one can find that the structure approaches its optimal performance for NV centers inside the lateral central areas of 40 nm radius and near diamond surface (< 20 nm) in the longitudinal direction. The traditional fabrication method adopts top-down processing without alignment (Fig. 1 (c)). The resulting distribution of NV centers inside the structure is random. In comparison, the self-aligned technique controls the diamond doping and fabrication both by a well-designed double-layer mask (Fig. 1 (d)). The top PMMA pattern acts as both a diamond doping mask to constrain the ion implantation region and a wet etching mask for isotropic wet etching of PMGI pattern. The bottom PMGI pattern constrains the self-alignment between the ion implantation region and the center of the resist.

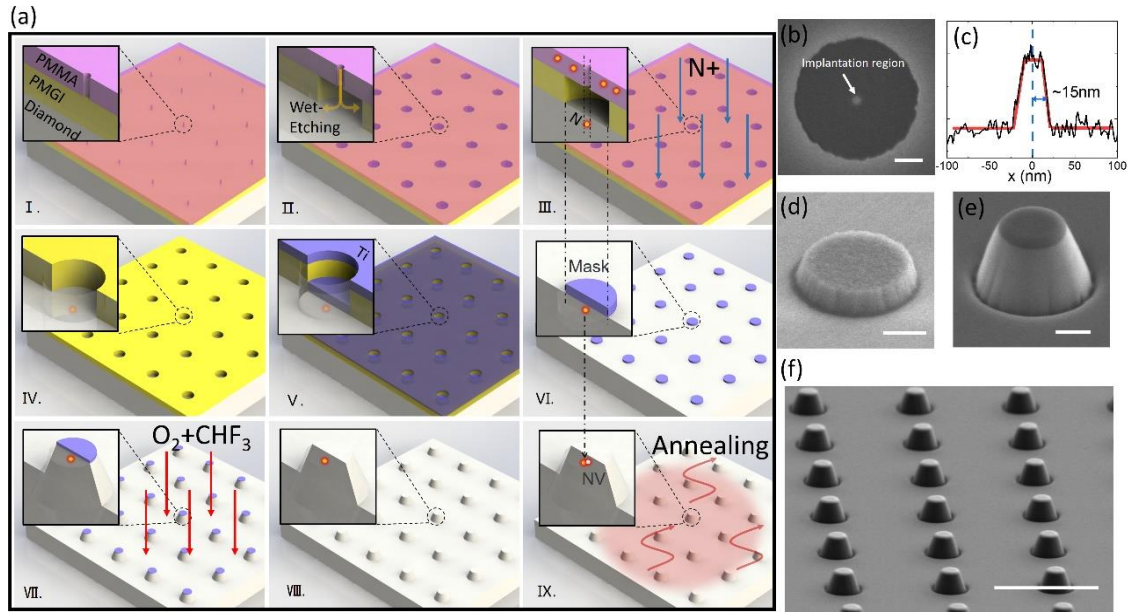

Figure 2 The fabrication process to realize self-aligned patterning and based sensor arrays (a) Schematic of fabrication steps. (I) Spin coating PMGI~270nm (yellow) + PMMA~210nm (pink) double-layer on diamond and making holes array in PMMA layer by electron beam lithography (EBL). (II) Isotropic wet etching of PMGI layer with TMAH (2.38%). (III) Doping nitrogen into diamond by ion implantation through holes array in PMMA layer. (IV) Removing the PMMA layer with acetone. (V) Coating titanium~100nm(blue) by electron beam evaporation (VI) liftoff processing in N-Methyl pyrrolidone （NMP）. (VII) Forming conical cylinder by $CHF_3+O_2$ reactive ion etching ($CHF_3$: $O_2$=5 sccm:30sccm). (VIII) Removing the Ti layer with buffered oxide etch (BOE). (IX) Converting nitrogen to NV center by annealing at 1000°C. (b) Verification of ion implanted region in (III) by metal deposition and SEM characterization (scale bar =100nm). The metal position in the center of the image represents the ion implantation region (See supplementary materials for details). (c) SEM imaging contrast shows that the radius of ion implantation region is ~15nm. (d) SEM image of a metal Ti mask in VI (scale bar =200nm). (e) SEM image of a conical cylinder by $CHF_3+O_2$ reactive ion etching in VII (scale bar =200nm). (f) SEM image of an array of conical cylinders(scale bar =2 μm).

Figure 2 (a) describes the whole process of self-alignment in detail. The double-layer mask is firstly prepared by the EBL process on PMMA, followed by isotropic wet etching of PMGI. The size of the PMGI mask is controlled by the wet etching time. A 5keV nitrogen ion implantation is then used to create shallow NV centers in the position constrained by the PMMA mask. The lateral straggling of ion caused by implantation can be ignored (about ~ 2 nm by SRIM simulation [48]). To characterize the implantation precision, we perform electron beam evaporation deposition of Ti on diamond through the prepared mask (See supplementary materials for details). As shown in Fig. 2(b)(c), SEM imaging of titanium pattern directly gives an ion implantation region radius of about 15 nm. Higher precision can be further obtained by depositing on the PMMA mask hole's sidewall[25,26]. After diamond doping, the double-layer mask is removed subsequently with a Ti spin coating in between, forming a Ti mask for the subsequent diamond etching. Fig. 2 (d) shows the Ti mask after the liftoff process. Fig. 2 (e) and (f) shows the fabricated nanopillar arrays etched with a fixed proportion of $CHF_3$ and $O_2$ mixed gas. This process also applies to preparing other nanostructures. By controlling the gas mixing, cylinder[6,49] and Parabolic reflector[50] can be fabricated similarly. Inverted Nanocones can be obtained by Faraday cage-assisted etching [10].

Integrated optical and optomechanical devices [51], such as the planar optical waveguide and ring resonator [52] [11], can be fabricated by changing the EBL pattern to line and ring.

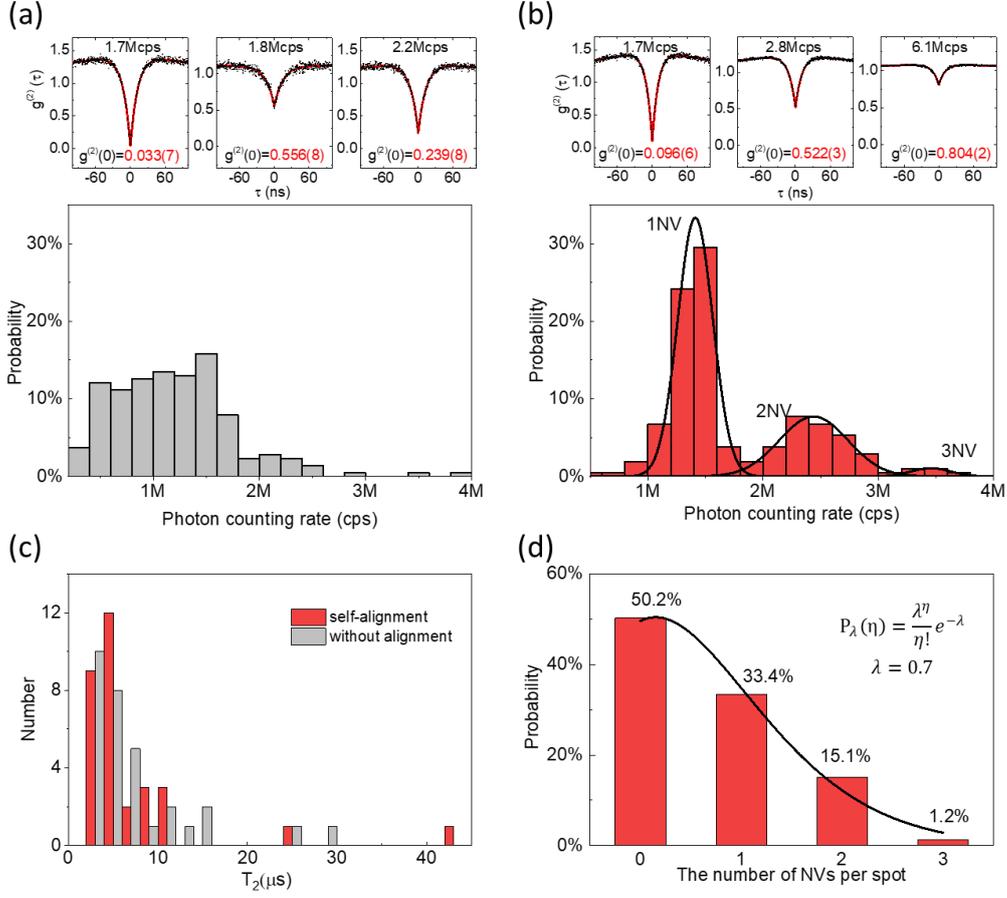

Figure 3 Comparison of the self-aligned sensors and non-aligned sensors prepared on (100)-oriented diamond. (a) Distribution of photon counting rate of NV sensors without alignment (at 800 μw excitation power). The insert shows the second order correlation function $g^{(2)}(\tau)$ (measured at ~40 μw excitation power) of the devices with different photon counting rates. (b) Distribution of photon counting rate of NV sensors by self-aligned (at 800 μw excitation power). The black line is the Gaussian fitting to the data envelope, and the number of NV centers in the photonic device can be distinguished by the envelope. The insert shows the second order correlation function $g^{(2)}(\tau)$ (measured at ~40 μw excitation power) of the devices with different photon counting rates. (c) Histogram of the $T_2$ coherence time, measured by Hahn echo pulse sequence. The red and grey bars show the NV center sensors fabricated by self-alignment and without alignment, respectively. (d) The distribution of the number of NV in self-aligned devices obtained from Fig3 (b). And fitted to a Poisson distribution.

**Performance of self-aligned sensors.** To characterize the self-aligned technique, we first compare NV sensors' performance with the same conical cylinder's shape, fabricated with and without self-alignment using (100)-oriented diamond. The radius of the self-aligned ion implantation mask is set at 40nm. 207 self-aligned sensors and 215 non-aligned sensors are investigated for the fluorescence intensity statics. As shown in Fig. 3 (a) (b), the non-aligned sensors show greatly varied photon counts due to randomly distributed NV centers, while the self-aligned sensors have high consistency and near-optimal photon counts. The second-order photon correlation measurement further verifies this. For non-aligned sensors with similar photon counts of about 1.7Mpcs, $g^{(2)}(0)$ shows distinct values indicating the presence of single and double NV centers ($g^{(2)}(0) < 0.5$ for single NV center and $g^{(2)} \sim 0.5$ for double NV centers). This observation suggests that it should be cautious about the actual number of NV centers and the corresponding photon counts due to the inconsistency in such non-aligned sensors. The criterion of the single-photon source ($g^{(2)}(0) < 0.5$) is no longer applicable here. A more complicated process like performing optically detected magnetic resonance in a gradient magnetic field is required for safe judgment [53]. In contrast, self-aligned sensors show high consistency, enabling the direct determination of the number of NV centers by photon counts. In this case the measured photon anti-bunching obeys the standard criterion ($g^{(2)}(0) = 0.096(6)$ for single NV centers of 1.7Mcps, $g^{(2)}(0) = 0.522(3)$ for double NV centers of 2.8 Mcps, and $g^{(2)}(0) = 0.804(2)$ for four NV centers of 6.1 Mcps). In Fig.3 (c), we count the number of NV centers in the aligned sensors according to the photon counts. The probability displays expected Poisson distribution due to ion implantation, further showing this high consistency. The proportion of single NV sensors has reached 33.4%($\lambda = 0.7$), close to the limit of 36.8% ($\lambda = 1$) under the optimal implantation dose condition. The coherence time of NV centers is also estimated by randomly measuring ~30 NV centers along the same direction. The statistical results (Fig.3 (d)) show similar coherence time distribution in both sensors.

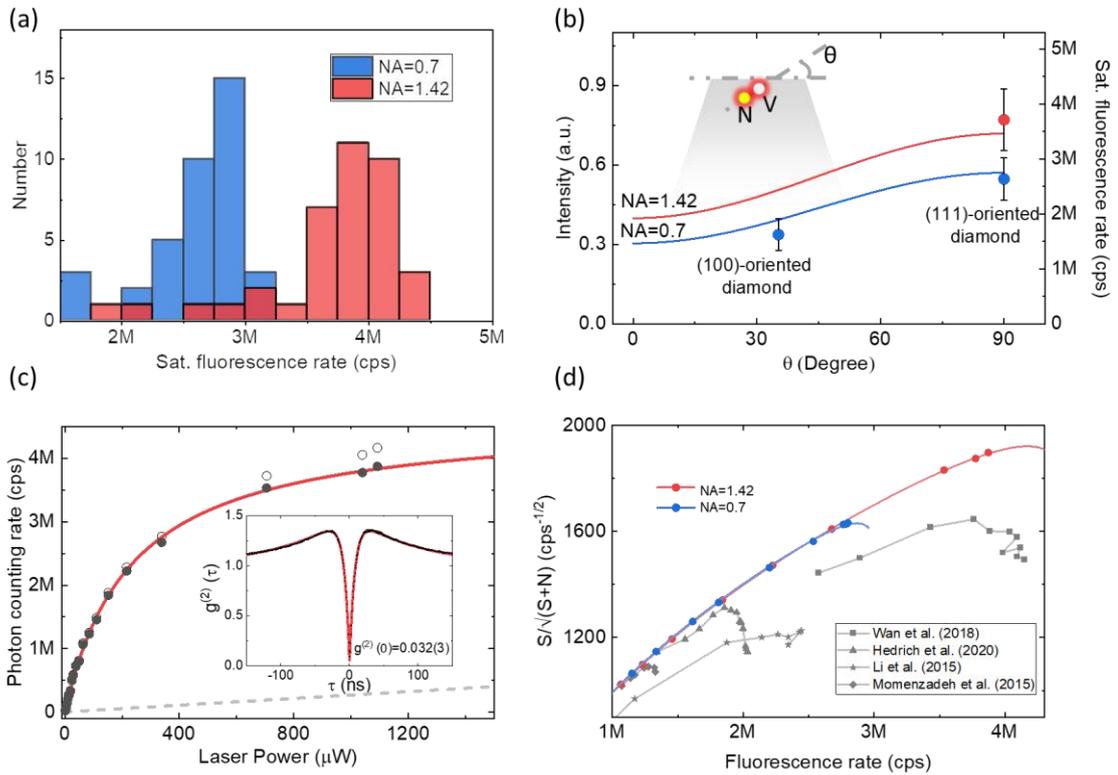

Figure 4 Performance of the nanopillar sensor arrays with direction control. (a) Histogram of the saturation fluorescence rate (background-subtracted) of NV centers. The blue bar and the red bar indicate the data measured with NA = 0.7 air/dry objective and NA = 1.42 oil-immersion objective respectively. (b) Comparison of simulated and experimental results for different θ (the angle between NV axis and diamond surface) with NA = 0.7 air/dry objective(blue)and NA = 1.42 oil-immersion objective(red). The solid line corresponding to the left axis is the result of simulation. The point corresponding to the right axis is the statistical mean of the experimental results, and the error bar is the statistical standard deviation. (c) Laser power dependent fluorescence count rate measurements for one

of the NV centers in (a) by NA = 1.42 oil-immersion objective. The hollow black circles are the photon counting rate data. The gray dash line is a linear fitting for the background obtained by $g^{(2)}(0)$ under different laser power. The bold black circles are the background-subtracted fluorescence rate. The red line is the fitted saturation curve and the saturation fluorescence rate is $4.65 \pm 0.04 \text{Mcps}$. The insert shows the $g^{(2)}(\tau)$ (measured at ~40 μw excitation power). (d) Fluorescence rate as a function of photon counting induced sensitivity. Our results are expressed in red (measured with NA = 1.42 oil-immersion objective) and blue (measured with NA = 0.7 air/dry objective objective), where the bold circle and solid line are obtained from the experimental data and saturation curve, respectively. The experimental results of NV center photonic structures reported are marked as a reference [7,8,9,50].

As another critical requirement, dipole direction control should be considered to achieve optimal performance devices. Because NV centers have deterministic [111] direction in the diamond lattice, we show this dipole direction control by adjusting the single-crystal diamond orientation. According to our FDTD simulation (Fig. 4 (b)), the (111)-oriented single crystal diamond, with one-fourth of NV centers perpendicular to the diamond surface, will have the best performance in photon counts. The corresponding experimental measurements are summarized in Fig.4(a). Benefiting from the self-aligned technique, the average saturated photon counts of NV centers is improved to 3.7±0.6 Mcps (with NA = 1.42 oil-immersion objective) and 2.6±0.4 Mcps (measured with NA = 0.7 air/dry objective). The photon counts of different directions obey the theoretical simulation (Fig. 4 (b)). The saturation fluorescence rate of the NV center shown in Fig. 4 (c) has reached 4.65 ± 0.04 Mcps, which reaches the optimal values in the best-reported diamond photonic structure [9]. Since NV sensors' magnetic sensitivity also accounts for the background of photon counts:

$$\delta B \propto 1/\frac{\sqrt{S+N}}{S}$$

Where $S$ is the fluorescence count rate of NV center and $N$ is background count rate. finally compare the photon counts induced sensitivity of $\frac{S}{\sqrt{S+N}}$ in all the reported sensors in Fig. 4. (d). Due to the high photon counts and low background, self-aligned sensors show the most heightened photon counts induced sensitivity of $1900 \text{ cps}^{-1/2}$.

## Conclusion

In summary, we propose a facile self-aligned patterning technique and demonstrate its power in fabricating high-performance diamond nanopillar sensors. The ~15 nm doping precision achieved is sufficient for fabricating other high-quality position sensitive emitter-photonic structures such as planar waveguides and ring resonators. The demonstrated precision can be further improved by sidewall deposition[25,26].

This technique is fully compatible with other diamond fabrication processes, like diamond surface treatment[54], high-temperature annealing[55,56], lattice charging [57], and n-type diamond [58-60], to further improve the creation yield, spin and optical properties of color centers in devices. The self-aligned patterning technique demonstrated here for NV centers could be extended to other similar solid-state systems such as silicon carbide, rare earth ions, etc. And this technique should facilitate the development of parallel quantum sensing and scalable information processing.

# Methods

**FDTD simulation.** The emission of fluorescence is associated with two orthogonal dipoles located in a plane perpendicular to the NV symmetry axis, which are marked by the red vectors in the inset of Fig. 1 (a) [3,61]. In FDTD simulation (Lumerical Solutions Ltd.), we take two incoherence electric dipoles($\lambda$ =637nm) that are orthogonal to each other and perpendicular to the NV axis to simulate the fluorescence emission of the NV center. The nanopillar model follows the actual structure, which was acquired through the scanning electron microscope: height~350nm, top diameter~380nm, and angle between sidewall and plane~69°. As shown in Fig. 1 (a), the monitor set below the nanopillar structure to get the dipoles emission far-field projection. And the far-field projection is recalculated based on Snell's law and the Fresnel equations to considering the reflection and refraction on the interface. The simulated fluorescence intensity is the integration of the far-field electric field intensity within the objective collection angle and normalized with source power.

**Device Fabrication.** The diamond used in experiment consisted of 50 μm thick (100)-oriented and (111)-oriented single-crystal diamond. The diamond was first cleaned in a boiling 1:1:1 nitric, perchloric and sulphuric acid bath to remove surface contamination.

For the sensors array with Self-align strategy (shown in Fig.2): PMGI ~ 270nm and PMMA ~ 210nm were spin coated on diamond, and hole arrays with 40 nm radius were fabricated on PMMA by 100keV electron beam lithography. The isotropic wet etching of PMGI pattern was carried out by 2.38% TMAH through holes in PMMA. The holes array in PMMA are used as ion implantation mask to dope ions into the center of PMGI pattern, and according to the simulation, the nitrogen ions with energy <20KeV would fully stopped by the 210nm thickness of the PMMA layer[48]. We implanted the sample with atomic nitrogen at an energy of 5 keV and dose of $5 \times 10^{11}/cm^2$. After ion implantation, the PMMA layer was removed by acetone and washed with isopropanol, and then dried by nitrogen. The pattern on PMGI is transferred to Ti masks by electron beam evaporation 100nm Ti and liftoff process in NMP. After that, the inductively coupled plasma (ICP) reactive-ion etching (RIE) system is used to etch the conical cylinder photonic structures, with $CHF_3$: $O_2$=5 sccm:30sccm mixed gas. after buffered oxide etching (BOE) of the residual Ti, the nitrogen is converted to NV center by annealing at 1000 degrees centigrade in vacuum.

For the sensors array without alignment: The maskless ion implantation is used with atomic nitrogen at an energy of 5 keV and dose of $6 \times 10^{10}/cm^2$, and the nitrogen is converted to NV center by annealing in 1000 degrees centigrade in vacuum. The Ti masks is directly made by electron beam lithography and liftoff process[26] and the same etching process as self-aligning sensor is used.

Before the test, both kinds of sensors are additional acid bath clean and annealing at 580 degrees centigrade in air for 20min.

**Experimental measurement set-up.** We used a home-built confocal scanning microscope with an excitation wavelength of 532 nm in the experiment. Two kinds of objective are used to excite and collect the fluorescence of NV center (Olympus, LUCPLFLN 60X NA=0.7; Olympus, PlanApoN 60x NA=1.42). We used diameter 75um pinhole on the platform to filter unfocused light. The fluorescence is filtered through a filter (RazorEdge, Long Pass 633nm) and eventually split into two equal-intensity beams by a D-shaped mirror and collected by two avalanche photodiode detectors (Excelitas, SPCM-AQRH-24). For the second-order autocorrelation function $g^{(2)}(\tau)$ measurement, the two detectors were directly connected to a time-correlated single-photon counting module (quTAG). And the data is normalized and fit by Three level system model: $g^{(2)}(\tau) = 1 + ce^{-|\tau|/\tau_b} - (1+c)e^{-|\tau|/\tau_a}$.

## Acknowledgements

The fabrication of diamond device was performed at the USTC Center for Micro and Nanoscale Research and Fabrication, and the authors particularly thank W. Liu, Y. Wei, X.W. Wang, C.L. Xin and H.F. Zuo for their assistance in EBL, deposition and etching process. The authors thank Prof. B. Cui for helpful discussion of PMGI etching. This work was supported by the National Key R&D Program of China (Grants No. 2018YFA0306600, 2017YFA0305000), the National Natural Science Foundation of China (Grants No. 81788101, 11775209, 12104447, 11804329), the CAS (Grants No. XDC07000000, GJJSTD20200001, QYZDYSSW-SLH004), the Anhui Initiative in Quantum Information Technologies (Grant No. AHY050000), the Fundamental Research Funds for the Central Universities, and USTC Research Funds of the Double First-Class Initiative (Grant No. YD2340002004), China Postdoctoral Science Foundation (Grant No. 2020M671858).

## Author contributions

J.D. and Y.W. proposed the idea and supervised the experiment. M.W. designed and performed the fabrication process. M.W. and H.S. collected and analyzed the experiment data. X.Y. and P.Y. assisted with ion implantation and annealing. M.W., P.Y., P.W., and F.S. constructed the experimental setup. Y.W., M.W. and J.D. wrote the paper. All authors discussed the results and commented on the manuscript.

# Supplementary materials

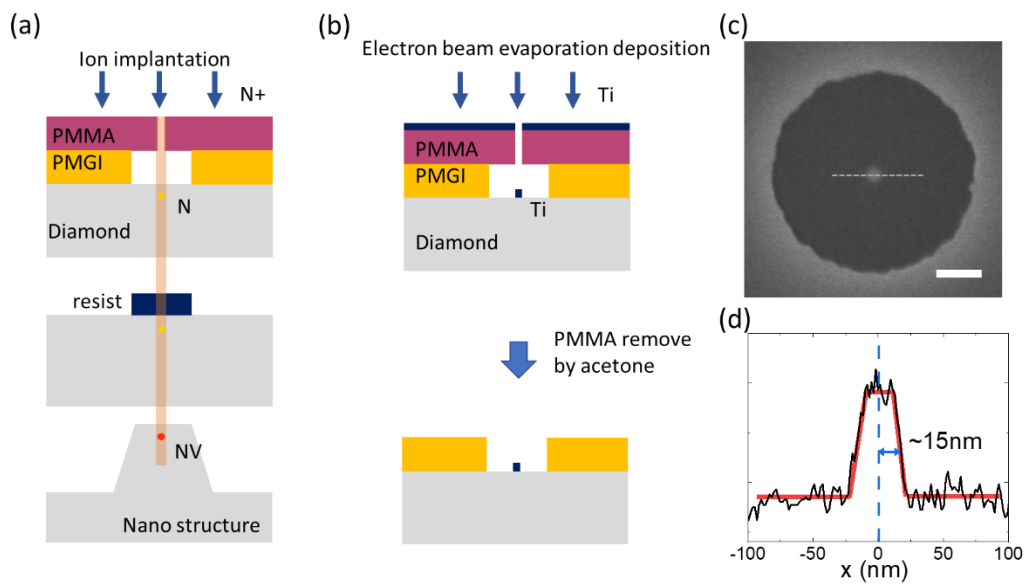

Figure S 1 **Characterization of the size and location of ion implanted region.** (a) Illustration of the fabrication process of photonic structures based on self-alignment. (b) Illustration of verification of ion implanted region by Ti deposition. Ti was deposited at 30 nm and PMMA was removed with acetone. (c) SEM image of the ion implantation region represented by Ti. (scale bar =100nm) (d) The contrast of the section shown by the dotted line in (c). The radius of the ion implantation region is ~15nm.

**Characterization of the size and location of ion implanted region**. In our self-aligned patterning technique, the PMMA and PMGI structures correspond to the ion implantation region and the etching mask, respectively. As shown in Fig. S1, the electron beam evaporation deposition has good directionality and is therefore used to verify the location and size of the ion implantation region. Removing PMMA with acetone and imaging PMGI structure and metal Ti by SEM can directly obtain the relative position between ion implantation and etching mask.

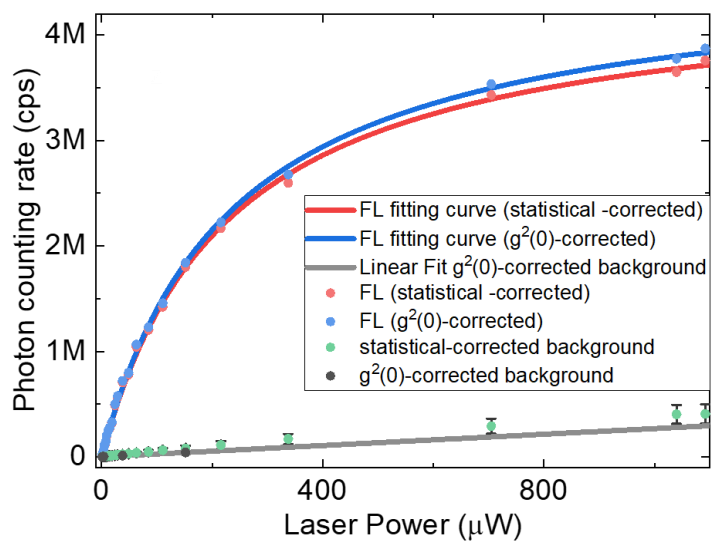

Figure S 2. The fluorescence rate background corrected by statistical correction and $g^2(0)$ correction.

The corrected fluorescence rate was fitted with $S(P) = S_{sat}/(1 + P_{sat}/P)$. The final fluorescence saturation rate are $S_{sat} = 4.49 \pm 0.04$ Mcps (saturation power $P_{sat} = 228 \pm 5\mu W$) and $S_{sat} = 4.65 \pm 0.04$ Mcps (saturation power $P_{sat} = 232 \pm 5\mu W$) for background subtraction with statistical correction method and g^2 (0) correction method, respectively.

**Background-subtracted method.** The statistical correction method and the $g^2(0)$ correction method are used to subtract the background in photon counting. The statistical correction method is to take the average of the photon counts of some photonic structures without NV center as the background counts, which is used for statistics of saturation fluorescence rate in Fig4(a)and(b). For the display of fluorescence of the NV center in Fig. 4 (c), we use $g^2(0)$ to correct the background of fluorescence. Background count rate $N = I * (1 - \sqrt{1 - g^2(0)})$, while $I$ is the photon counting rate. And the background as a function of laser power is obtained by linear fitting. As shown in Fig. S2, the difference between the two correction methods can be ignored for the NV center in Fig. 4(c).